\documentclass{article}
\usepackage{eurosym}
\usepackage{amssymb}
\usepackage{amsmath}
\usepackage{cite}
\usepackage{graphicx}
\usepackage{epsf}
\usepackage{lscape}

\begin{document}

\title{Similarity solutions for two-phase fluids models}
\author{Andronikos Paliathanasis\thanks{%
Email: anpaliat@phys.uoa.gr} \\
{\ \ \textit{Institute of Systems Science, Durban University of Technology }}%
\\
{\ \textit{PO Box 1334, Durban 4000, Republic of South Africa}\ } \\
{\textit{Instituto de Ciencias F\'{\i}sicas y Matem\'{a}ticas,}}\\
{\ \textit{Universidad Austral de Chile, Valdivia, Chile}}}
\maketitle

\begin{abstract}
The algebraic properties of drift-flux two-phase fluids models without
gravitational and wall friction forces are studied. More precisely, for the
two fluids we consider equation of states of polytropic gases. We perform a
classification scheme of the unknown parameters of the model such that to
determine all the possible admitted Lie symmetries. We find that in the most
general case the dynamical system of hyperbolic equations is invariant under
the action of a four dimensional Lie algebra, while the larger number of
admitted Lie symmetries is six. For each admitted Lie algebra the
one-dimensional optimal system is derived which is applied for the
determination of all the unique similarity transformations which lead to
similarity solutions. Our results are compared with that of previous studies
from where we see that most of the solutions presented in this study have
not found before in the literature.

\bigskip

\end{abstract}

\section{Introduction}

\label{sec1}

A systematic approach for the determination of exact and analytical
solutions of nonlinear differential equations is Lie's theory \cite%
{lie1,lie2,lie3}. The novelty of Lie's consideration is that someone
investigate the invariance properties of a given differential equation under
the action of infinitesimal transformations in order to write the
differential equation into an equivalent form through algebraic
representation \cite{ibra,Bluman,Stephani,olver}.\ The generator of the
infinitesimal transformation which keeps a differential equation invariant
is called Lie symmetry of the right type. A main property for the admitted
Lie symmetries for a differential equation is that they form an algebra
known as Lie algebra.

The existence of a large-dimensional Lie algebra of Lie symmetries for a
differential equation enables one to solve the differential equation by
means of repeated reduction of order with the use of similarity
transformations, or by means of the determination of a sufficient number of
first integrals \cite{m1,m2}. In particular if an $n$th-order differential
equation admit at least $n$ Lie point symmetries with a solvable algebra,
one knows that the system is reducible to quadratures and the system is
integrable \cite{ls}. Thus, the absence of the latter property does not
immediately obviate the possibility of integrability. For instance, the
application of a similarity transformation reduce the differential equation
into a new differential equation which may has different algebraic
properties and additional (new) symmetries to follow, \cite{ab1}. On the
other hand, there is a zoology on the nature of symmetries which means that
because an equation does not admit Lie point symmetries it does not means
that it is not invariant under the action of other kind transformations
which follows from generalized forms of the definition of symmetries for
instance see \cite{s1,s2,s5} and references therein. Another important
application of the symmetries is that they are applied for the
classification of equations and establish classes and families of equations
which has as common feature the common admitted Lie algebra \cite%
{Mubar1,Mubar2,Mubar3,Sharp1,Sharp2,Sharp3}.

There are various applications in the literature on the application of Lie
symmetries for the study of nonlinear differential equations. Some results
on ordinary differential equations are presented in \cite{od1,od3}, while
there are various studies and for elliptic partial differential equations,
for instance see \cite{el3,el4,el5} and references therein. As far as the
application of Lie symmetries on hyperbolic partial differential equations
is concerned, there are various of studies in all areas of applied
mathematics \cite{hp1,hp2,hp3,hp4}. In theory of fluid dynamics Lie
symmetries have played an important role on the determination of exact and
analytic solutions. The complete symmetry classification of the
shallow-water equations with or without a gravitational field performed in
\cite{sw3,swr06}, while the case with a Coriolis force in \cite{an2,an3}.
The case with various class of bottoms investigated in \cite{mm1,mm2,mm3}
while for other studies on shallow-water equations we refer the reader in
\cite{mel1,sw4,sw9} and references therein. \ A detailed discussion of \ the
symmetry approach on mechanical theories of continuous media is presented in
\cite{ds1}. The authors by studying some systems of physical interests
demonstrate the application of Lie invariants for the derivation of
similarity solutions.

Another important system of hyperbolic equations where Lie point symmetries
have been used for the determination of new solutions is the two-phase flow
model \cite{bir1,bir2}.~The two-phase fluids model describes the evolution
of two fluids with different phases, such as liquid and gas, in a tube. The
two-phase models have many physical applications from oil extraction,
underground water, nuclear reactor and many others \cite{rt2}. For some
important results on two-phase flow models we refer the reader in \cite%
{tw3,tw4,tw7,tw8,tw9,zd1,zd2} and references therein. The exact solution of
the Riemann problem for the drift-flux equations for a two-phase flow system
was studied in \cite{zd3} while numerical simulations of wave propagation in
compressible two-phase flow were presented in \cite{zd4}.

In \cite{bir1,bir2} Lie's theory applied for the simplest two-phase flow
model where there is not any mass transfer from the one fluid to the other
while the pressure and the energy density of the two fluids is given by an
polytropic equation of state as it is given by Lane-Emden equation. In this
work we revise the results of \cite{bir1,bir2} for a non-flip drift flux
model of multi-phase flow defined \cite{banda1}. More specifically, we find
new solutions which have not been presented before, for linear and nonlinear
equation of state parameters for the two fluids. More precisely, we derive
the admitted Lie point symmetries of the two-phase fluids model and for the
admitted Lie symmetries we calculate the commutators and the Adjoint
representation. By using these results we are able to determine the
one-dimensional optimal system such that to perform all the independent
similarity transformations. We find that the similarity solutions are
expressed either by closed-form expressions or by quadratures. There are not
many known exact solutions in two-phase models, see for instance \cite%
{f1,f2,f3}, hence the analysis presented here is important for the
determination of exact solutions. The plan of the paper is as follows.

The main mathematical properties and definitions on the symmetries of
differential equations are presented in Section \ref{sec2}. The main results
of this study are presented in Section \ref{sec3}. The drift-flux two-phase
fluids model is given where we determine the admitted Lie point symmetries.
The unknown parameters of the problem are the two polytropic exponents $%
\gamma _{1},~\gamma _{2}$ of the two fluids. For arbitrary values of the
polytropic exponents such that $\gamma _{i}\geq 1,$ \ the set of three
hyperbolic equations admit four Lie point symmetries, while two additional
Lie symmetries exist when $\gamma _{1}=\gamma _{2}=\gamma $ which means that
in the latter case the dynamical system is invariant under a sixth
dimensional Lie algebra. For each different admitted Lie algebra we
calculate the commutators and the Adjoint representation. By using these
results we are able to derive the one-dimensional optimal system, necessary
for the derivation of all the independent similarity solutions. On the
derivation of the similarity solutions emphasis is given in the two cases in
which $\gamma _{1}=\gamma _{2}=2$ and $\gamma _{1}=\gamma _{2}=1$. Finally,
in \ref{sec4} we discuss our results and we draw our conclusions.

\section{Lie symmetries of differential equations}

\label{sec2}

In this Section we present the basic properties and definitions of the Lie
point symmetries. Consider the system of differential equations $H^{A}\left(
x^{i},\Phi ^{A},\Phi _{,i}^{A},...\right) =0$ where $x^{i}$ denotes the
indepedent variables and $\Phi ^{A}$ are the independent variables while a $%
\Phi _{,i}^{A}$ indicates derivative with respect to the variable $x^{i}$,
that is, $\Phi _{,i}^{A}=\frac{\partial }{\partial x^{i}}\Phi ^{A}$.

We proceed by assuming the infinitesimal one-parameter point transformation
of the form
\begin{equation}
\bar{x}^{i}=x^{i}\left( x^{j},\Phi ^{B};\varepsilon \right) ,~~\bar{\Phi}%
^{A}=\Phi ^{A}\left( x^{j},\Phi ^{B};\varepsilon \right) ,  \label{de.03}
\end{equation}%
where $\varepsilon ~$is the infinitesimal parameter. Point transformation (%
\ref{de.03}) connects two different points $P\left( x^{j},\Phi ^{B}\right)
\rightarrow Q\left( \bar{x}^{j},\bar{\Phi}^{B},\varepsilon \right) $, while
at these two points the system of differential equations is defined $%
H^{A}\left( P\right) $ and $\bar{H}^{A}\left( Q\right) $.

We shall say that the system $H^{A}$ remains invariant under the action of
the one-parameter transformation if and only if $\bar{H}^{A}=H^{A}$, that is
\cite{ibra,Bluman,Stephani,olver}%
\begin{equation}
\lim_{\varepsilon \rightarrow 0}\frac{\bar{H}^{A}\left( \bar{y}^{i},\bar{u}%
^{A},...;\varepsilon \right) -H^{A}\left( y^{i},u^{A},...\right) }{%
\varepsilon }=0.  \label{ls.05}
\end{equation}%
Which means that the solutions $\Phi ^{A}~$of the system $H^{A}$ at the two
different points, i.e. $\Phi ^{A}\left( P\right) $ and $\Phi ^{A}\left(
Q\right) ,$ are related through the point transformation (\ref{de.03}).

By definition, expression (\ref{ls.05}) is the Lie derivative of $H^{A}$
along the vector field $X$ of the one-parameter point transformation (\ref%
{de.03}), in which $X$ is defined as
\begin{equation*}
X=\frac{\partial \bar{x}^{i}}{\partial \varepsilon }\partial _{i}+\frac{%
\partial \bar{\Phi}}{\partial \varepsilon }\partial _{A}.
\end{equation*}%
Hence, an equivalent form of the symmetry condition is%
\begin{equation}
\mathcal{L}_{X}\left( H^{A}\right) =0,  \label{ls.05a}
\end{equation}%
where~$\mathcal{L}$ denotes the Lie derivative with respect to the vector
field $X^{\left[ n\right] }.$ Vector field $X^{\left[ n\right] }$ the $n$%
th-extension of generator $X~$of the transformation (\ref{de.03}) in the jet
space $\left\{ x^{i},\Phi ^{A},\Phi _{,i}^{A},\Phi _{,ij}^{A},...\right\} $
where $H^{A}$ is defined. The vector field $X^{\left[ n\right] }$ is
calculated by the generator of the point transformation $X$ as follows
\begin{equation}
X^{\left[ n\right] }=X+\eta ^{\left[ 1\right] }\partial _{\Phi
_{i}^{A}}+...+\eta ^{\left[ n\right] }\partial _{\Phi
_{i_{i}i_{j}...i_{n}}^{A}},  \label{ls.06}
\end{equation}%
where the new terms are%
\begin{equation}
\eta ^{\left[ n\right] }=D_{i}\eta ^{\left[ n-1\right]
}-u_{i_{1}i_{2}...i_{n-1}}D_{i}\left( \frac{\partial \bar{x}^{j}}{\partial
\varepsilon }\right) ~,~i\succeq 1~,~\eta ^{\left[ 0\right] }=\left( \frac{%
\partial \bar{\Phi}^{A}}{\partial \varepsilon }\right) .  \label{de.08}
\end{equation}

The Lie symmetries for a given differential equation form a Lie algebra. Lie
symmetries can be used by different ways\ \cite{Bluman} in order to study a
differential equation. However, their direct application is on the
determination of the so-called similarity solutions. The steps which we
follow to determine a similarity solution is based on the determination and
application of the Lie invariant functions.

In order to determine all the possible independent solutions of a given
dynamical system we should derive the one-dimensional optimal system. Let
the $n$-dimensional Lie algebra $G_{n}$ with elements $%
X_{1},~X_{2},~...~X_{n}$ admitted by the system $H^{A}$. The two vector
fields \cite{olver}
\begin{equation}
Z=\sum\limits_{i=1}^{n}a_{i}X_{i}~,~W=\sum\limits_{i=1}^{n}b_{i}X_{i}~,~%
\text{\ }a_{i},~b_{i}\text{ are constants.}  \label{sw.04}
\end{equation}%
are equivalent if and only if%
\begin{equation}
\mathbf{W}=Ad\left( \exp \left( \varepsilon _{i}X_{i}\right) \right) \mathbf{%
Z}  \label{sw.05}
\end{equation}%
or%
\begin{equation}
W=cZ~,~c=const.  \label{sw.06}
\end{equation}%
where the operator $Ad\left( \exp \left( \varepsilon X_{i}\right) \right)
X_{j}=X_{j}-\varepsilon \left[ X_{i},X_{j}\right] +\frac{1}{2}\varepsilon
^{2}\left[ X_{i},\left[ X_{i},X_{j}\right] \right] +...~$is called the
adjoint representation.$~$

\section{Two-phase flow model}

\label{sec3}

Let us assume a mixture fluid stratified flow in a pipe, where the mixture
is consisting by two phases of the same fluid, i.e. liquid $\left( \rho
_{l}\right) $\ and gas $\left( \rho _{g}\right) $. Furthermore, each phase
moves on the pipe at a local section with average velocity $u_{l}$\ and $%
u_{g}$\ . Hence, the continuous equations for the two fluids in the
one-dimensional read \cite{thorne}%
\begin{eqnarray}
\left( \rho _{l}a_{l}\right) _{,t}+\left( \rho _{l}a_{l}u_{l}\right) _{,x}
&=&\Gamma _{l}a_{l},  \label{ss.01} \\
\left( \rho _{g}a_{g}\right) _{,t}+\left( \rho _{g}a_{g}u_{g}\right) _{,x}
&=&\Gamma _{g}a_{g},  \label{ss.02}
\end{eqnarray}%
where $\Gamma _{l},~\Gamma _{g}$\ are the particle creation terms, where for
a closed system hold $\Gamma _{l}=-\Gamma _{g}$.~Furthermore, $a_{l},~a_{g}~$%
are the volume fractions for the two fluids with $a_{l}+a_{g}=1$. We
continue our analysis by assume the simplest model without particle creation
term, or any interaction between the two fluids, while we omit any
gravitational effect.

Hence, the momentum equations for the fluids reads \cite{thorne}%
\begin{equation}
\left( \rho _{l}a_{l}u_{l}\right) _{,t}+\left( \rho
_{l}a_{l}u_{l}^{2}+a_{l}p_{l}\right) _{,x}-p_{l}^{i}\left( a_{l}\right)
_{,x}=0,  \label{ss.03}
\end{equation}%
\begin{equation}
\left( \rho _{g}a_{g}u_{g}\right) _{,t}+\left( \rho
_{g}a_{g}u_{g}^{2}+a_{g}p_{g}\right) _{,x}-p_{g}^{i}\left( a_{g}\right)
_{,x}=0.  \label{ss.04}
\end{equation}%
in which $p_{l},~p_{g}$\ are the pressure terms for the two fluids and $%
p_{l}^{i},~p_{g}^{i}$\ describe the interfacial pressures on the gas--liquid
interface on the side of the liquid and of the gas respectively.

Hence, from the momentum equations (\ref{ss.03}), (\ref{ss.04}) it follows%
\begin{eqnarray}
0 &=&\left( \rho _{l}a_{l}u_{l}+\rho _{g}a_{g}u_{g}\right) _{,t}+\left( \rho
_{l}a_{l}u_{l}^{2}+\rho _{g}a_{g}u_{g}^{2}\right) _{,x}+  \notag \\
&&+a_{l}p_{l,x}+a_{g}p_{g,x}+\left( p_{l}-p_{l}^{i}\right) \left(
a_{l}\right) _{,x}+\left( p_{g}-p_{g}^{i}\right) \left( a_{g}\right) _{,x}
\label{ss.05}
\end{eqnarray}

Therefore, with the use of the new variables $u_{r}=u_{g}-u_{l}$\ and $u=%
\frac{\rho _{g}a_{g}u_{g}+\rho _{l}a_{l}u_{l}}{\rho _{g}a_{g}+\rho _{l}a_{l}}%
+u_{r},$\ and by considering that for the drift model that $u_{r}=0$, we end
with the finally system \cite{bir2}

\begin{eqnarray}
\rho _{1,t}+\left( \rho _{1}u\right) _{,x} &=&0,  \label{tp.01} \\
\rho _{2,t}+\left( \rho _{2}u\right) _{,x} &=&0,  \label{tp.02} \\
\left( \left( \rho _{1}+\rho _{2}\right) u\right) _{,t}+\left( \left( \rho
_{1}+\rho _{2}\right) u^{2}+p_{1}+p_{2}\right) _{,x} &=&0.  \label{tp.03}
\end{eqnarray}%
where $\rho _{\left( 1,2\right) }\left( t,x\right) =a_{\left( l,g\right)
}\rho _{\left( l,g\right) }\left( t,x\right) $\ are the energy density of
the fluid of phase and $p_{\left( 1,2\right) }\left( t,x\right) =a_{\left(
l,g\right) }p_{\left( l,g\right) }\left( t,x\right) $\ denote the pressure
of the fluids. Recall that for the drift flux model, by definition, the
pressure in the gas and the liquid surfaces are equal for the same
cross-sectional area \cite{thorne}.

Finally, for the fluids we assume polytropic equation of state parameters of
the form $p_{\left( 1,2\right) }=\kappa _{\left( 1,2\right) }\rho ^{\gamma
_{\left( 1,2\right) }}$\ with $\gamma _{\left( 1,2\right) }\geq 1$.\

\subsection{Arbitrary polytropic exponents}

For the system (\ref{tp.01})-(\ref{tp.03}) we apply Lie's theory and we
determine the Lie point symmetries which are \cite{bir2}%
\begin{equation*}
X_{1}=\partial _{t}~,~X_{2}=\partial _{x}~,~X_{3}=t\partial _{t}+x\partial
_{x}\text{ and }X_{4}=t\partial _{x}+\partial _{u}
\end{equation*}%
for arbitrary values of the polytropic exponents $\gamma _{1},~\gamma _{2}$.
The commutators of the admitted Lie symmetries by the system (\ref{tp.01})-(%
\ref{tp.03}) and the Adjoint representation of the Lie algebra consisted by
the elements $\left\{ X_{1},X_{2},X_{3},X_{4}\right\} $are presented in
Tables \ref{tab1} and \ref{tab2}. The admitted Lie algebra is a solvable
four-dimensional Lie algebra classified as $A_{4,6}$ in the Patera and
Winternitz classification scheme \cite{Sharp1}.

From Tables \ref{tab1} and \ref{tab2} we infer that the one-dimensional
optimal system is consisted by the elements%
\begin{eqnarray*}
&&\left\{ X_{1}\right\} ~,~\left\{ X_{2}\right\} ~,~\left\{ X_{3}\right\}
~,~\left\{ X_{4}\right\} ~,~ \\
&&\left\{ X_{1}+\beta X_{2}\right\} ~,~\left\{ X_{3}+\beta X_{4}\right\}
~,\left\{ X_{1}+\beta X_{4}\right\} .
\end{eqnarray*}

\begin{table}[tbp] \centering%
\caption{Commutator table for the Lie point symmetries of the two-phase flow
system (\ref{tp.01})-(\ref{tp.03})}%
\begin{tabular}{ccccc}
\hline\hline
$\left[ X_{I},X_{J}\right] $ & $\mathbf{X}_{1}$ & $\mathbf{X}_{2}$ & $%
\mathbf{X}_{3}$ & $\mathbf{X}_{4}$ \\ \hline
$\mathbf{X}_{1}$ & $0$ & $0$ & $X_{1}$ & $X_{2}$ \\
$\mathbf{X}_{2}$ & $0$ & $0$ & $X_{2}$ & $0$ \\
$\mathbf{X}_{3}$ & $-X_{1}$ & $-X_{2}$ & $0$ & \thinspace $0$ \\
$\mathbf{X}_{4}$ & $-X_{2}$ & $0$ & $0$ & $0$ \\ \hline\hline
\end{tabular}%
\label{tab1}%
\end{table}%

\begin{table}[tbp] \centering%
\caption{Adjoint representation for the Lie point symmetries of the two-phase flow
system (\ref{tp.01})-(\ref{tp.03})}%
\begin{tabular}{ccccc}
\hline\hline
$Ad\left( e^{\left( \varepsilon \mathbf{X}_{i}\right) }\right) \mathbf{X}%
_{j} $ & $\mathbf{X}_{1}$ & $\mathbf{X}_{2}$ & $\mathbf{X}_{3}$ & $\mathbf{X}%
_{4}$ \\ \hline
$\mathbf{X}_{1}$ & $X_{1}$ & $X_{2}$ & $-\varepsilon X_{1}+X_{3}$ & $%
-\varepsilon X_{2}+X_{4}$ \\
$\mathbf{X}_{2}$ & $X_{1}$ & $X_{2}$ & $-\varepsilon X_{2}+X_{3}$ & $X_{4}$
\\
$\mathbf{X}_{3}$ & $e^{\varepsilon }X_{1}$ & $e^{\varepsilon }X_{2}$ & $%
X_{3} $ & $X_{4}$ \\
$\mathbf{X}_{4}$ & $X_{1}+\varepsilon X_{2}$ & $X_{2}$ & $X_{3}$ & $X_{4}$
\\ \hline\hline
\end{tabular}%
\label{tab2}%
\end{table}%

We proceed with the application of the Lie symmetries such that to find the
invariant transformations where the solution of the system (\ref{tp.01})-(%
\ref{tp.03}) is expressed by the solution of an ordinary differential
equation.

\subsubsection{Reduction with $X_{1}$}

Application of the Lie point symmetry $X_{1}$ gives stationary solutions,
that is, $\rho _{i}=\rho _{i}\left( x\right) ,~u=u\left( x\right) $ where
the reduced system is
\begin{equation}
\left( \rho _{1}u\right) _{,x}=0~,~\left( \rho _{2}u\right) _{,x}=0~,\left(
\left( \rho _{1}+\rho _{2}\right) u^{2}+\kappa _{1}\rho _{1}^{\gamma
_{1}}+\kappa _{2}\rho _{2}^{\gamma _{2}}\right) _{,x}=0,  \label{tp.04}
\end{equation}%
which means $\rho _{1}u=\rho _{10},~\rho _{2}u=\rho _{20}$ and
\begin{equation}
\left( \rho _{10}+\rho _{20}\right) u+\kappa _{1}\left( \rho _{10}\right)
^{\gamma _{1}}u^{-\gamma _{1}}+\kappa _{2}\left( \rho _{20}\right) ^{\gamma
_{2}}u^{-\gamma _{2}}=u_{10}.  \label{tp.05}
\end{equation}%
Thus, it follows, $u\left( x\right) =const$.

\subsubsection{Reduction with $X_{2}$}

From the Lie point symmetry $X_{2}$ the stationary similarity transformation
are $\rho _{i}=\rho _{i}\left( t\right) ,~u=u\left( t\right) $, where%
\begin{equation}
\rho _{1,t}=0~,~\rho _{2,t}=0~,~\left( \left( \rho _{1}+\rho _{2}\right)
u\right) _{,t}=0,  \label{tp.06}
\end{equation}%
that is, $\rho _{i}=\rho _{i0}$ and $u=u_{0}$.

\subsubsection{Reduction with $X_{3}$}

From the scaling symmetry $X_{3}$ it follows the similarity transformation~$%
\rho _{i}=\rho _{i}\left( w\right) ,~u=u\left( w\right) $ where the new
independent variable is defined as $w=\frac{x}{t}$. The reduced system is%
\begin{eqnarray}
w\rho _{1,w}-\left( \rho _{1}u\right) _{,w} &=&0,  \label{tp.07} \\
w\rho _{2,w}-\left( \rho _{2}u\right) _{,w} &=&0,  \label{tp.08} \\
w\left( \left( \rho _{1}+\rho _{2}\right) u\right) _{,w}+\left( \left( \rho
_{1}+\rho _{2}\right) u^{2}+p_{1}+p_{2}\right) _{,w} &=&0.  \label{tp.09}
\end{eqnarray}

From (\ref{tp.07}) and (\ref{tp.08}) it follows%
\begin{equation}
\rho _{i}\left( w\right) =\rho _{i0}\exp \left( -\int \frac{u_{,w}}{u-w}%
dw\right) ,  \label{tp.10}
\end{equation}%
where by replacing in (\ref{tp.09}) we end with the equation%
\begin{equation}
U_{,ww}\left( \left( U_{,w}\right) ^{-1}\left( \rho _{01}+\rho _{02}\right)
\left( U_{0}-U\right) ^{2}-\kappa _{1}\left( \rho _{01}\right) ^{\gamma
_{1}}\left( U_{,w}\right) ^{\gamma _{1}}-\kappa _{1}\left( \rho _{02}\right)
^{\gamma _{1}}\left( U_{,w}\right) ^{\gamma _{2}}\right) =0,  \label{tp.11}
\end{equation}%
in which we have defined $u\left( w\right) =w-\left( U_{0}-U\left( w\right)
\right) \left( U_{,w}\right) ^{-1}$.

Hence, function $U\left( w\right) $ is given by the non-static solution of
the first-order differential equation%
\begin{equation}
\left( U_{,w}\right) ^{-1}\left( \rho _{01}+\rho _{02}\right) \left(
U_{0}-U\right) ^{2}-\kappa _{1}\left( \rho _{01}\right) ^{\gamma _{1}}\left(
U_{,w}\right) ^{\gamma _{1}}-\kappa _{1}\left( \rho _{02}\right) ^{\gamma
_{1}}\left( U_{,w}\right) ^{\gamma _{2}}=0.  \label{tp.12}
\end{equation}

The latter equation is solved by quadratures, however there are some
closed-form solutions for specific values of the polytropic exponents $%
\gamma _{i}$. Indeed, for $\gamma _{i}=1$ the closed-form solution is
\begin{equation}
U_{\pm }\left( w\right) =U_{0}-U_{1}e^{\pm \lambda w}  \label{tp.13}
\end{equation}%
where now $\lambda =\sqrt{\frac{\rho _{01}+\rho _{02}}{\kappa _{1}\rho
_{0}+\kappa _{2}\rho _{02}}}$, thus the similarity solution is
\begin{equation}
u\left( w\right) =w-\frac{1}{\lambda }~,~\rho _{i}\left( w\right) =\rho
_{i0}e^{\lambda w}.  \label{tp.14}
\end{equation}%
This is a new\ close-form solution which has not found before.

\subsubsection{Reduction with $X_{4}$}

The similarity transformation given by the symmetry vector $X_{4}$ is $\rho
_{i}=\rho _{i}\left( t\right) $,~$u=\frac{x}{t}+v\left( t\right) $ where
\begin{eqnarray}
t\rho _{1,t}+\rho _{1} &=&0,  \label{tp.15} \\
t\rho _{2,t}+\rho _{2} &=&0,  \label{tp.16} \\
\left( t\left( \rho _{1,t}+\rho _{2,t}\right) +\left( \rho _{1}+\rho
_{2}\right) \right) x+\left( \rho _{1}v+\rho _{2}v\right) _{,t}+2t\left(
\rho _{1}+\rho _{2}\right) v &=&0,  \label{tp.17}
\end{eqnarray}%
from where it follows the exact solution%
\begin{equation}
\rho _{i}\left( t\right) =\rho _{i0}t^{-1}\text{ and }v\left( t,x\right) =%
\frac{\left( x+u_{0}\right) }{t}\text{. }  \label{tp.18}
\end{equation}

\subsubsection{Travellin waves $X_{1}+\protect\beta X_{2}$}

From the Lie symmetry $X_{1}+\beta X_{2}$ it follows $\rho _{i}=\rho
_{i}\left( w\right) ,~u=u\left( w\right) $ where $w=x-\beta t$ while the
reduced system is
\begin{eqnarray}
\beta \rho _{1,w}-\left( \rho _{1}u\right) _{,w} &=&0,  \label{tp.19} \\
\beta \rho _{2,w}-\left( \rho _{2}u\right) _{,w} &=&0,  \label{tp.20} \\
\left( \left( \rho _{1}+\rho _{2}\right) u\right) _{,w}-\beta \left( \left(
\rho _{1}+\rho _{2}\right) u^{2}+p_{1}+p_{2}\right) _{,w} &=&0,
\label{tp.21}
\end{eqnarray}%
which provides the constant solution $\rho _{i}=\rho _{i0}$ and $u=u_{0}$,
or for $u=\beta $ it follows
\begin{equation}
\rho _{1}\left( w\right) =\left( \rho _{10}-\frac{\kappa _{2}}{\kappa _{1}}%
\rho _{2}\left( w\right) \right) ^{\frac{1}{\gamma _{1}}}  \label{tp.22}
\end{equation}%
where $\rho _{2}\left( w\right) $ is an arbitrary function.

Recall that in order the latter solution to be physically accepted $\rho
_{2}\left( w\right) $ should be defined such that $\rho _{1},\rho _{2}$ to
be real positive functions. That is a more general solution that the one
found before in \cite{bir2}. The only solution which was found in \cite{bir2}
provides that one of the $\rho _{i}\left( w\right) $ will be negative, or
one of the pressure $p_{i}\left( \rho _{i}\left( w\right) \right) $ will be
negative. That is not physically accepted, since there are not known
physically fluids with negative energy density or negative pressure.
Although that kind of fluids are used in theoretical astrophysics as toy
models, they have not observed yet.

\subsubsection{Reduction with $X_{3}+\protect\beta X_{4}$}

Application of the symmetry vector $X_{3}+\beta X_{4}$ gives the similarity
transformation $\rho _{i}=\rho _{i}\left( w\right) ,~u=\beta \ln t+v\left(
w\right) $ with $w=x-\beta \ln t$. The reduced system provides%
\begin{equation}
\rho _{i}\left( w\right) =\rho _{i0}V_{,w}~,~v\left( w\right) =\left(
w+\beta \right) +\frac{V_{0}-V\left( w\right) }{V_{,w}}  \label{tp.23}
\end{equation}%
where $V\left( w\right) $ satisfies the second-order ordinary differential
equation
\begin{equation}
0=\left( \rho _{10}+\rho _{20}\right) \left( \beta \left( V_{,w}\right)
^{3}-V_{,ww}\left( V-V_{0}\right) ^{2}\right) +V_{,w}V_{,ww}\left( \kappa
_{1}\left( \rho _{10}V_{,w}\right) ^{\gamma _{1}}+\kappa _{2}\left( \rho
_{20}V_{,w}\right) ^{\gamma _{2}}\right) ,  \label{tp.24}
\end{equation}

Equation (\ref{tp.24}) is an autonomous equation, hence we can define the
new dependent variable $\phi =V_{,w}$ and the new independent variable $z=V$.

Therefore, it follows%
\begin{equation*}
0=\left( \rho _{10}+\rho _{20}\right) \left( \beta \left( \phi \left(
z\right) \right) ^{3}-\phi \left( z\right) \phi _{,z}\left( z-V_{0}\right)
^{2}\right) +\phi _{,z}\left( \phi \left( z\right) \right) ^{2}\left( \kappa
_{1}\left( \rho _{10}\phi \left( z\right) \right) ^{\gamma _{1}}+\kappa
_{2}\left( \rho _{20}\phi \left( z\right) \right) ^{\gamma _{2}}\right) ,
\end{equation*}%
which can be solved by quadratures.

\subsubsection{Reduction with $X_{1}+\protect\beta X_{4}$}

From the symmetry vector $X_{1}+\beta X_{4}$ it follows $\rho _{i}=\rho
_{i}\left( w\right) $,~$u=\beta t+V\left( w\right) $ where now $w=x-\frac{%
\beta }{2}t^{2}$. The reduced system is%
\begin{equation}
\left( \rho _{1}v\right) _{,w}=0~,~\left( \rho _{2}v\right) _{,w}=0~,
\end{equation}%
\begin{equation}
\beta \left( \rho _{1}+\rho _{2}\right) +\left( \left( \rho _{1}+\rho
_{2}\right) u^{2}+p_{1}+p_{2}\right) _{,w}=0
\end{equation}%
that is $\rho _{1}\left( w\right) =\rho _{10}v^{-1},~~\rho _{2}\left(
w\right) =\rho _{20}v^{-1}$. Hence $v\left( w\right) $ is given by the
equation
\begin{equation}
\left( \rho _{10}+\rho _{20}\right) w+v_{0}+\frac{1}{2}\left( \rho
_{10}+\rho _{20}\right) v^{2}-\frac{\kappa _{1}\left( \rho _{10}\right)
^{\gamma _{1}}}{1-\gamma _{1}}v^{1-\gamma _{1}}-\frac{\kappa _{1}\left( \rho
_{20}\right) ^{\gamma _{2}}}{1-\gamma _{2}}v^{1-\gamma _{2}}=0
\end{equation}%
for $\gamma _{i}\neq 1$, or
\begin{equation}
\left( \rho _{10}+\rho _{20}\right) w+v_{0}+\frac{1}{2}\left( \rho
_{10}+\rho _{20}\right) v^{2}-\kappa _{1}\rho _{10}\ln v-\frac{\kappa
_{1}\left( \rho _{20}\right) ^{\gamma _{2}}}{1-\gamma _{2}}v^{1-\gamma
_{2}}=0.
\end{equation}%
for $\gamma _{1}=1,~\gamma _{2}\neq 1$ or%
\begin{equation}
\left( \rho _{10}+\rho _{20}\right) w+v_{0}+\frac{1}{2}\left( \rho
_{10}+\rho _{20}\right) v^{2}-\kappa _{1}\rho _{10}\ln v^{\kappa _{1}\rho
_{10}+\kappa _{2}\rho _{29}}=0.
\end{equation}%
for $\gamma _{i}=1$. This is a new solutions which has not found before in
the literature.

Nevertheless for specific values of the polytropic exponents the dynamical
system (\ref{tp.01})-(\ref{tp.03}) admits additional symmetries.

\subsection{Polytropic exponents $\protect\gamma _{1}=\protect\gamma _{2}=%
\protect\gamma ~,~\protect\gamma \neq 1$}

Consider now the case where the polytropic exponents $\gamma _{i}$ are
equal, that is, $\gamma _{1}=\gamma _{2}=\gamma $. In such scenario the Lie
point symmetries of the two-phase model (\ref{tp.01})-(\ref{tp.03}) are the
vector fields $X_{1},~X_{2},~X_{3},~X_{4},$ plus the additional symmetries%
\begin{equation*}
~X_{5}=\frac{2}{\gamma -1}\left( \rho _{1}\partial _{\rho _{1}}+\rho
_{2}\partial _{\rho _{2}}+u\partial u-t\partial _{t}\right)
~,~X_{6}=f_{1}\left( \rho _{1},\rho _{2};\kappa _{1},\kappa _{2},\gamma
\right) \partial _{\rho _{1}}+f_{2}\left( \rho _{1},\rho _{2};\kappa
_{1},\kappa _{2},\gamma \right) \partial _{\rho _{2}}.
\end{equation*}

Vector field $X_{5}$ is a scaling symmetry, while $X_{6}$ is a rotation
symmetry which indicates the invariance of the two-phase model if $\left(
\rho _{1},\rho _{2}\right) \rightarrow \left( \bar{\rho}_{1},\bar{\rho}%
_{2}\right) $. However, functions $f_{i}\left( \rho _{1},\rho _{2};\kappa
_{1},\kappa _{2},\gamma \right) $ are not expressed always into closed-form
expressions. Thus for $\gamma =2$ it follows%
\begin{eqnarray}
f_{1}\left( \rho _{1},\rho _{2};\kappa _{1},\kappa _{2},\gamma \right) &=&%
\frac{\kappa _{1}\rho _{1}^{2}-2\kappa _{2}\rho _{1}\rho _{2}-\kappa
_{2}\rho _{2}^{2}}{\kappa _{1}\rho _{1}-\kappa _{2}\rho _{2}}, \\
f_{2}\left( \rho _{1},\rho _{2};\kappa _{1},\kappa _{2},\gamma \right) &=&%
\frac{\kappa _{1}\rho _{1}^{2}+2\kappa _{1}\rho _{1}\rho _{2}-\kappa
_{2}\rho _{2}^{2}}{\kappa _{1}\rho _{1}-\kappa _{2}\rho _{2}}.
\end{eqnarray}%
For arbitrary polytropic index $\gamma $ we found that $f_{i}=f_{i}\left(
\frac{\rho _{2}}{f_{1}}\right) \rho _{i}$, which indicates that $\left[
X_{5},X_{6}\right] =0$.

Symmetry vector $X_{6}$ has not been derived before in the literature and it
is a new symmetry. In Table \ref{tab3} the commutators of the admitted Lie
symmetries is presented while in Table \ref{tab4} the corresponding adjoint
representation is derived.

From Tables \ref{tab3} and \ref{tab4} the one-dimensional optimal system is
calculated consisted by the vector fields%
\begin{eqnarray*}
&&\left\{ X_{1}\right\} ~,~\left\{ X_{2}\right\} ~,~\left\{ X_{3}\right\}
~,~\left\{ X_{4}\right\} ~,~\left\{ X_{5}\right\} ~,~\left\{ X_{6}\right\} ~,
\\
&&\left\{ X_{1}+\beta X_{2}\right\} ~,~\left\{ X_{3}+\beta X_{4}\right\}
~,\left\{ X_{1}+\beta X_{4}\right\} ~, \\
&&\left\{ X_{2}+\beta X_{5}\right\} ~,~\left\{ X_{3}+\beta X_{5}\right\}
~,~\left\{ X_{1}+\beta X_{6}\right\} ~, \\
&&\left\{ X_{2}+\beta X_{6}\right\} ~,~\left\{ X_{3}+\beta X_{6}\right\}
~,\left\{ X_{4}+\beta X_{6}\right\} ~, \\
&&\left\{ X_{5}+\beta X_{6}\right\} ~~,\left\{ X_{2}+\beta X_{5}+\delta
X_{6}\right\} ~,~\left\{ X_{3}+\beta X_{5}+\delta X_{6}\right\} .
\end{eqnarray*}%
We continue our analysis with the new reductions, while for simplicity on
our calculations and on the presentation we select the polytropic index $%
\gamma =2$.

\begin{table}[tbp] \centering%
\caption{Commutator table for the Lie point symmetries of the two-phase flow
system (\ref{tp.01})-(\ref{tp.03}) for equal indices $\gamma_{i}$}%
\begin{tabular}{ccccccc}
\hline\hline
$\left[ X_{I},X_{J}\right] $ & $\mathbf{X}_{1}$ & $\mathbf{X}_{2}$ & $%
\mathbf{X}_{3}$ & $\mathbf{X}_{4}$ & $\mathbf{X}_{5}$ & $\mathbf{X}_{6}$ \\
\hline
$\mathbf{X}_{1}$ & $0$ & $0$ & $X_{1}$ & $X_{2}$ & $-X_{1}$ & $0$ \\
$\mathbf{X}_{2}$ & $0$ & $0$ & $X_{2}$ & $0$ & $0$ & $0$ \\
$\mathbf{X}_{3}$ & $-X_{1}$ & $-X_{2}$ & $0$ & \thinspace $0$ & \thinspace $%
0 $ & $0$ \\
$\mathbf{X}_{4}$ & $-X_{2}$ & $0$ & $0$ & $0$ & $X_{4}$ & $0$ \\
$\mathbf{X}_{5}$ & $X_{1}$ & $0$ & $0$ & $-X_{4}$ & $0$ & $0$ \\
$\mathbf{X}_{6}$ & $0$ & $0$ & $0$ & $0$ & $0$ & $0$ \\ \hline\hline
\end{tabular}%
\label{tab3}%
\end{table}%

\begin{table}[tbp] \centering%
\caption{Adjoint representation for the Lie point symmetries of the two-phase flow
system (\ref{tp.01})-(\ref{tp.03}) for equal indices $\gamma_{i}$}%
\begin{tabular}{ccccccc}
\hline\hline
$Ad\left( e^{\left( \varepsilon \mathbf{X}_{i}\right) }\right) \mathbf{X}%
_{j} $ & $\mathbf{X}_{1}$ & $\mathbf{X}_{2}$ & $\mathbf{X}_{3}$ & $\mathbf{X}%
_{4}$ & $\mathbf{X}_{5}$ & $\mathbf{X}_{6}$ \\ \hline
$\mathbf{X}_{1}$ & $X_{1}$ & $X_{2}$ & $-\varepsilon X_{1}+X_{3}$ & $%
-\varepsilon X_{2}+X_{4}$ & $\varepsilon X_{1}+X_{5}$ & $\mathbf{X}_{6}$ \\
$\mathbf{X}_{2}$ & $X_{1}$ & $X_{2}$ & $-\varepsilon X_{2}+X_{3}$ & $X_{4}$
& $X_{5}$ & $\mathbf{X}_{6}$ \\
$\mathbf{X}_{3}$ & $e^{\varepsilon }X_{1}$ & $e^{\varepsilon }X_{2}$ & $%
X_{3} $ & $X_{4}$ & $X_{5}$ & $\mathbf{X}_{6}$ \\
$\mathbf{X}_{4}$ & $X_{1}+\varepsilon X_{2}$ & $X_{2}$ & $X_{3}$ & $X_{4}$ &
$-\varepsilon X_{4}+X_{5}$ & $\mathbf{X}_{6}$ \\
$\mathbf{X}_{5}$ & $e^{\varepsilon }X_{1}$ & $X_{2}$ & $X_{3}$ & $%
e^{\varepsilon }X_{4}$ & $X_{5}$ & $\mathbf{X}_{6}$ \\
$\mathbf{X}_{6}$ & $X_{1}$ & $X_{2}$ & $X_{3}$ & $X_{4}$ & $X_{5}$ & $%
\mathbf{X}_{6}$ \\ \hline\hline
\end{tabular}%
\label{tab4}%
\end{table}%

\subsubsection{Reduction with $X_{5}$}

Application of the symmetry vector $X_{5}$ provides the similarity
transformation $\rho _{i}=\rho _{i}\left( x\right) t^{-2}$ and $u=u\left(
x\right) t^{-2}$, where
\begin{equation}
\rho _{i}\left( x\right) =\rho _{i0}\exp \left( \int \frac{2-u_{,x}}{u}%
dx\right)
\end{equation}%
where $u=\left( 2v\left( x\right) +v_{0}\right) \left( v_{,x}\right) ^{-1}$
and $v\left( x\right) $ is given by the differential equation%
\begin{equation}
v_{,xx}\left( \left( \kappa _{1}\rho _{10}^{2}+\kappa _{2}\rho
_{20}^{2}\right) \left( v_{,x}\right) ^{3}-\left( 2v+v_{0}\right) ^{2}\left(
\rho _{10}+\rho _{20}\right) \right) +\left( v_{,x}\right) ^{2}\left( \rho
_{10}+\rho _{20}\right) \left( 2v+v_{0}\right) =0,
\end{equation}%
the later equation can be integrated by quadratures.

\subsubsection{Reduction with $X_{2}+\protect\beta X_{5}$}

From the Lie symmetry vector~$X_{2}+\beta X_{5}$ we find the similarity
transformation $\rho _{i}=\rho _{i}\left( w\right) t^{-2}~,~u=\bar{u}\left(
w\right) t^{-1}$ where~$w=x+\frac{1}{\beta }\ln t$. The reduced system
provides%
\begin{equation}
\rho _{i}=\rho _{i0}\exp \left( \beta \int \frac{2-\bar{u}_{,w}}{1+\beta
\bar{u}_{,w}}dw\right)
\end{equation}%
where
\begin{equation}
\bar{u}=-\frac{1}{\beta }+\left( 2v\left( w\right) +v_{0}\right) \left(
v_{,w}\right) ^{-1}
\end{equation}%
where now $v\left( x\right) $ is given by the second-order differential
equation%
\begin{eqnarray}
0 &=&\beta v_{,ww}\left( 2\left( \kappa _{1}\rho _{10}^{2}+\kappa _{2}\rho
_{20}^{2}\right) \left( v_{,w}\right) ^{3}-\left( 2v+v_{0}\right) ^{2}\left(
\rho _{10}+\rho _{20}\right) \right)  \notag \\
&&+\left( v_{,w}\right) ^{2}\left( \rho _{10}+\rho _{20}\right) \left( \beta
\left( 2v+v_{0}\right) +v_{,w}\right) .
\end{eqnarray}%
which again can be solved by quadratures.

\subsubsection{Reduction with $X_{3}+\protect\beta X_{5}$}

The Lie symmetry vector $\left\{ X_{3}+\beta X_{5}\right\} $ provides the
similarity transformation $\rho _{i}=\rho _{i}\left( w\right) t^{-\frac{%
2\beta }{\beta -1}},~u=\bar{u}\left( w\right) t^{-\frac{\beta }{\beta -1}}$
with new independent variable $w=xt^{\frac{1}{\beta -1}}$, while the reduced
system provides%
\begin{equation*}
\rho _{i}\left( w\right) =\rho _{i0}v_{x}~,~\bar{u}=-\frac{1}{\beta }+\left(
2v\left( w\right) +v_{0}\right) \left( v_{,w}\right) ^{-1}
\end{equation*}%
where $v\left( w\right) $ satisfies the second-order differential equation
\begin{eqnarray}
0 &=&v_{,ww}\left( \left( \rho _{10}+\rho _{20}\right) \left( v\left( 2\beta
+1\right) -v_{0}\left( 1-\beta \right) \right) -2\left( \kappa _{1}\rho
_{10}^{2}+\kappa _{2}\rho _{20}^{2}\right) \left( v_{,w}\right) ^{3}\left(
1-\beta \right) ^{2}\right) +  \notag \\
&&-w\left( v_{,w}\right) ^{3}\beta \left( \rho _{10}+\rho _{20}\right)
-\left( \rho _{10}+\rho _{20}\right) \beta \left( v_{,w}\right) ^{2}\left(
v\left( 2\beta +1\right) -v_{0}\left( 1-\beta \right) \right)
\end{eqnarray}%
which can be solved by quadratures.

\subsubsection{Reduction with $X_{1}+\protect\beta X_{6}$}

In order to proceed with the reduction with the use of the rest of the
symmetry vectors we define the new variables%
\begin{equation}
\rho \left( t,x\right) =\rho _{1}\left( t,x\right) +\rho _{2}\left(
t,x\right) ,  \label{tp.25}
\end{equation}%
\begin{equation}
p\left( t,x\right) =\kappa _{1}\rho _{1}^{2}\left( t,x\right) +\kappa
_{2}\rho _{2}^{2}\left( t,x\right) ,  \label{tp.26}
\end{equation}%
where now system (\ref{tp.01})-(\ref{tp.03}) becomes%
\begin{eqnarray}
\rho _{,t}+\left( \rho u\right) _{,x} &=&0,  \label{tp.27} \\
p_{,t}+\left( pu\right) _{,x}+pu_{,x} &=&0,  \label{tp.28} \\
\left( \rho u\right) _{,t}+\left( \rho u^{2}+p\right) _{,x} &=&0.
\label{tp.29}
\end{eqnarray}%
In the new coordinates, symmetries $X_{5,}~X_{6}$ are written $X_{5}=-2\rho
\partial _{\rho }+u\partial _{u}-t\partial _{t}$ and $X_{6}=\rho \partial
_{\rho }+p\partial _{p}\,$.

Thus, application of the Lie symmetry vector $X_{1}+\beta X_{6}$ gives
\begin{equation}
\rho =e^{t}\rho \left( x\right) ~,~p=e^{t}p\left( x\right) ~,~u=u\left(
x\right)
\end{equation}%
while the reduced system is%
\begin{eqnarray}
\rho +\left( \rho u\right) _{,x} &=&0, \\
p+\left( pu\right) _{,x}+pu_{,x} &=&0, \\
\rho u+\left( \rho u^{2}+p\right) _{,x} &=&0
\end{eqnarray}%
which gives the solution%
\begin{equation}
\rho \left( x\right) =\frac{2pu_{,x}+\beta p}{u^{2}u_{,x}}~,~p=p_{0}\exp
\left( -\int \frac{\beta +2u_{,x}}{u}dx\right)
\end{equation}%
where $u\left( x\right) $ is given by the polynomial%
\begin{equation}
u_{0}\beta u^{4}-2\beta u-x-\beta x_{0}=0.
\end{equation}%
This is also a new solution for the two-phase fluids model. As also all the
following similarity solutions are new and they have not been calculated
before.

\subsubsection{Reduction with $X_{2}+\protect\beta X_{6}$}

The similarity transformation which follows from $X_{2}+\beta X_{6}$ is $%
\rho =e^{\beta x}\bar{\rho}\left( t\right) ~,~p=e^{\beta x}\bar{p}\left(
t\right) $ and $u=u\left( t\right) $ while system (\ref{tp.27})-(\ref{tp.29}%
) becomes%
\begin{eqnarray}
\bar{\rho}_{t}+\beta \bar{\rho}u &=&0, \\
\bar{p}_{t}+\beta \bar{p}u &=&0, \\
\left( \bar{\rho}u\right) _{,t}+\beta \bar{\rho}u^{2}+\beta \bar{p} &=&0,
\end{eqnarray}%
from where we infer the physically accepted solution
\begin{eqnarray}
u\left( t\right) &=&u_{0}t+u_{1},~ \\
\bar{p}\left( t\right) &=&p_{0}\exp \left( -\frac{\beta }{2}\lambda
_{0}t^{2}-\beta \lambda _{1}t\right) ~,~ \\
\bar{\rho}\left( t\right) &=&-\frac{\beta p_{0}}{\lambda _{0}}\exp \left( -%
\frac{\beta }{2}\lambda _{0}t^{2}-\beta \lambda _{1}t\right) ~.
\end{eqnarray}

\subsubsection{Reduction with $X_{3}+\protect\beta X_{6}$}

From the symmetry vector $X_{3}+\beta X_{6}$ we find $\rho =t^{\beta }\bar{%
\rho}\left( w\right) ,~p=t^{\beta }\bar{p}\left( w\right) $ and $u=u\left(
w\right) $ with $w=\frac{x}{t}$. Hence, from (\ref{tp.27})-(\ref{tp.29}) we
find the solution%
\begin{equation}
\bar{\rho}\left( w\right) =\rho _{0}\exp \left( \frac{\beta +u_{,w}}{w-u}%
\right) ~,~\bar{p}=p_{1}\exp \left( \int \frac{\beta -2u_{w}}{w-u}\right) ,
\end{equation}%
where $u\left( w\right) =w+v\left( w\right) $, while $v\left( w\right) $ is
given by the integral $\int^{v\left( w\right) }\left( f\left( s\right)
\right) ^{-1}ds-\left( w-w_{0}\right) =0$, where $f\left( s\right) $ is a
solution of the polynomial equation%
\begin{equation}
-s^{-2-\frac{3}{2}\beta }\left( f\left( s\right) +1\right) ^{-1-\frac{3}{4}%
\beta }\left( 3f\left( s\right) +1\right) ^{\frac{\beta }{4}}\left( 2f\left(
s\right) +\beta +2\right) ^{\frac{\beta }{2}+1}+p_{0}=0.
\end{equation}

\subsubsection{Reduction with $X_{4}+\protect\beta X_{6}$}

Application of the symmetry vector $X_{4}+\beta X_{6}$ gives the
transformation $\rho =e^{\beta \frac{x}{t}}\bar{\rho}\left( t\right)
~,~p=e^{\beta \frac{x}{t}}\bar{p}\left( t\right) $~and $u=\frac{x}{t}%
+v\left( t\right) $, where now the reduced system is derived%
\begin{eqnarray}
t\bar{\rho}_{,t}+\rho \left( 1+\beta v\right) &=&0, \\
t\bar{p}_{,t}+p\left( 2+\beta v\right) &=&0, \\
t\left( \bar{\rho}v\right) _{,t}+\rho \left( \beta v^{2}+2v\right) +\beta
\bar{p} &=&0,
\end{eqnarray}%
that is%
\begin{eqnarray}
v\left( t\right) &=&\frac{u_{0}}{t}+\frac{u_{1}}{t}\ln t~, \\
\bar{p}\left( t\right) &=&p_{0}t^{\frac{\beta }{t}-2}\exp \left( \frac{\beta
}{t}\left( u_{0}+u_{1}\right) \right) ~, \\
\bar{\rho}\left( t\right) &=&-\beta \frac{p_{0}}{u_{1}}t^{\frac{\beta }{t}%
-1}\exp \left( \frac{\beta }{t}\left( u_{0}+u_{1}\right) \right) .
\end{eqnarray}%
Thus, in order the solution to be physically accepted, $p_{0}>0$ and $\beta
u_{1}<0$.

\subsubsection{Reduction with $X_{5}+\protect\beta X_{6}$}

The Lie symmetry vector $X_{5}+\beta X_{6}$ provides $\rho =t^{2-\beta }\bar{%
\rho}\left( x\right) ,~p=t^{-\beta }\bar{p}\left( x\right) $ and $u=t^{-1}%
\bar{u}\left( x\right) $, while the reduced system is%
\begin{eqnarray}
\left( 2-\beta \right) \bar{\rho}+\left( \bar{\rho}\bar{u}\right) _{,x} &=&0,
\\
-\beta \bar{p}+\left( \bar{p}u\right) _{,x}+\bar{p}\bar{u}_{,x} &=&0, \\
\left( 1-\beta \right) \bar{\rho}\bar{u}+\left( \bar{\rho}\bar{u}^{2}+\bar{p}%
\right) _{,x} &=&0.
\end{eqnarray}%
\qquad that is,%
\begin{eqnarray}
\bar{\rho}\left( x\right) &=&\rho _{0}\exp \left( -\int \frac{2-\beta +\bar{u%
}_{,x}}{\bar{u}}dx\right) ~, \\
\bar{p}\left( x\right) &=&p_{0}\exp \left( -\int \frac{2\bar{u}_{,x}-\beta }{%
\bar{u}}dx\right) ,
\end{eqnarray}%
where $\bar{u}\left( w\right) $ is given by the expression%
\begin{equation}
\int^{\bar{u}\left( x\right) }\left( g\left( s\right) \right) ^{-1}ds=\left(
x-x_{0}\right) ,
\end{equation}%
where
\begin{equation*}
-2\left( sg\left( s\right) -1\right) ^{\frac{4-3\beta }{\beta }}\left(
3g\left( s\right) -2\right) ^{-\frac{2}{\beta }\left( 2-\beta \right)
}\left( 2g\left( s\right) -\beta \right) +g_{0}=0\text{.}
\end{equation*}

\subsubsection{Reduction with $X_{2}+\protect\beta X_{5}+\protect\delta %
X_{6} $}

From $X_{2}+\beta X_{5}+\delta X_{6}$ it follows $\rho =t^{2-\frac{\delta }{%
\beta }}\bar{\rho}\left( w\right) ,~p=t^{-\frac{\delta }{\beta }}\bar{p}%
\left( w\right) ,~u=t^{-1}\bar{u}\left( w\right) $, where $w=x+\frac{1}{%
\beta }\ln t$. Thus, system (\ref{tp.27})-(\ref{tp.29}) is reduced
\begin{eqnarray}
\bar{\rho}_{,w}+\left( 2\beta -\delta \right) \bar{\rho}+\beta \left( \bar{%
\rho}\bar{u}\right) _{,w} &=&0, \\
\bar{p}_{,w}-\delta \bar{p}+\beta \left( \left( \bar{p}\bar{u}\right) _{,w}+%
\bar{p}\bar{u}_{,w}\right) &=&0, \\
\left( \bar{\rho}\bar{u}\right) _{,w}+\left( \beta -\delta \right) \bar{\rho}%
\bar{u}+\beta \left( \bar{\rho}\bar{u}^{2}+\bar{p}\right) _{,w} &=&0.
\end{eqnarray}%
The latter system can be integrated further and its solution is expressed in
terms of quadratures.

\subsubsection{Reduction with $X_{3}+\protect\beta X_{5}+\protect\delta %
X_{6} $}

Similarly, from the symmetry vector $X_{3}+\beta X_{5}+\delta X_{6}$ it
follows $\rho =t^{\frac{2\beta -\delta }{\beta -1}}\bar{\rho}\left( w\right)
,~p=t^{-\frac{\delta }{\beta -1}}\bar{p}\left( w\right) ,~u=t^{-\frac{\beta
}{\beta -1}}\bar{u}\left( w\right) $ with $w=xt^{\frac{1}{\beta -1}}$.
Therefore, the reduced system is
\begin{eqnarray}
\bar{\rho}_{,w}+\left( 2\beta -\delta \right) \bar{\rho}+\left( \beta
-1\right) \left( \bar{\rho}\bar{u}\right) _{,w} &=&0, \\
\bar{p}_{,w}-\delta \bar{p}+\left( \beta -1\right) \left( \left( \bar{p}\bar{%
u}\right) _{,w}+p\bar{u}_{,w}\right) &=&0, \\
\left( \bar{\rho}\bar{u}\right) _{,w}+\left( \beta -\delta \right) \bar{\rho}%
\bar{u}+\left( \beta -1\right) \left( \bar{\rho}\bar{u}^{2}+\bar{p}\right)
_{,w} &=&0.
\end{eqnarray}%
which again can be integrated by quadratures.

We continue our analysis with the special case of the polytropic exponents $%
\gamma _{1}=\gamma _{2}=1$.

\subsection{Polytropic exponents $\protect\gamma _{1}=\protect\gamma _{2}=1$}

In the special case where the two polytropic exponents are equal with one,
that is, $\gamma _{1}=\gamma _{2}=1$. Similarly with the previous case we
define the new dependent variables%
\begin{equation}
\rho \left( t,x\right) =\rho _{1}\left( t,x\right) +\rho _{2}\left(
t,x\right) ,  \label{tp.30}
\end{equation}%
\begin{equation}
p\left( t,x\right) =\kappa _{1}\rho _{1}\left( t,x\right) +\kappa _{2}\rho
_{2}\left( t,x\right) ,  \label{tp.31}
\end{equation}%
where the dynamical system (\ref{tp.01})-(\ref{tp.03}) takes the following
form%
\begin{eqnarray}
\rho _{,t}+\left( \rho u\right) _{,x} &=&0,  \label{tp.32} \\
p_{,t}+\left( pu\right) _{,x} &=&0,  \label{tp.33} \\
\left( \rho u\right) _{,t}+\left( \rho u^{2}+p\right) _{,x} &=&0.
\label{tp.34}
\end{eqnarray}

The set of equations (\ref{tp.32})-(\ref{tp.34}) admits a six dimensional
Lie algebra consisted by the symmetry vectors $\left\{
X_{1},X_{2},X_{3},X_{4},X_{5},X_{6}\right\} $. Thus, there is not any
different between the symmetry classification of the two cases $\gamma =1$
and $\gamma \neq 1$, as it was found before in \cite{bir2}. We proceed with
the application of the symmetry vectors for the derivation of similarity
solutions.

\subsubsection{Reduction with $X_{5}$}

Application of the symmetry vector $X_{5}$ gives $\rho =t^{-2}\bar{\rho}%
\left( x\right) ,~p=p\left( x\right) $ and $u=t^{-1}\bar{u}\left( x\right) $
where the reduced system is
\begin{eqnarray}
\left( \bar{\rho}\bar{u}\right) _{,x}+2\bar{\rho} &=&0, \\
\left( p\bar{u}\right) _{,x} &=&0, \\
\bar{\rho}\bar{u}+\left( \bar{\rho}\bar{u}^{2}+p\right) _{,x} &=&0,
\end{eqnarray}%
from where we find $\bar{\rho}=v_{,x}$,~$p=p_{0}u^{-1}$ and $\bar{u}=\frac{%
2\left( v_{0}-v\right) }{v_{,x}}$, where $v\left( x\right) $ satisfies the
differential equation%
\begin{equation}
6\rho _{0}\left( v-v_{0}\right) +\frac{p_{0}}{2\left( v-v_{0}\right) ^{2}}%
\left( \left( v_{0}-v\right) v_{,xx}+v_{,x}^{2}\right) -\frac{4\rho _{0}}{%
v_{,x}^{2}}\left( v-v_{0}\right) ^{2}v_{,xx}=0,
\end{equation}%
which can be solved by quadratures.

\subsubsection{Reduction with $X_{2}+\protect\beta X_{5}$}

The vector field $X_{2}+\beta X_{5}$ provides $\rho =t^{2}\rho \left(
w\right) ,~p=p\left( w\right) $ and $u=t^{-1}u\left( w\right) $ where now
the independent variable is defined as $w=x+\frac{1}{\beta }\ln t$. Hence,
by replacing in (\ref{tp.32})-(\ref{tp.34}) we find%
\begin{equation}
\rho \left( w\right) =\rho _{0}v_{,w}~,~p\left( w\right) =p_{0}\left(
1+\beta u\right) ^{-1}~,~u\left( w\right) =-\frac{1}{\beta }+\frac{2\left(
v_{0}-v\right) }{v_{,x}}
\end{equation}%
where $v\left( x\right) $ satisfies the second-order ode%
\begin{eqnarray}
0 &=&v_{xx}\left( v_{0}-v\right) \left( 8\beta \rho _{0}\left(
v-v_{0}\right) ^{3}+p_{0}\left( v_{,x}\right) ^{2}\right) +p_{0}\left(
v_{,x}\right) ^{4}+  \notag \\
&&+2\rho _{0}\left( v-v_{0}\right) ^{2}\left( v_{x}\right) ^{3}+12\beta \rho
_{0}\left( v_{,x}\right) ^{2}\left( v-v_{0}\right) ^{3}
\end{eqnarray}%
which can be solved by quadratures.

\subsubsection{Reduction with $X_{3}+\protect\beta X_{5}$}

From the vector field $X_{3}+\beta X_{5}$ it follows $\rho =t^{\frac{2\beta
}{\beta -1}}\bar{\rho}\left( w\right) ,~p=p\left( w\right) $ and $u=t^{-%
\frac{\beta }{\beta -1}}\bar{u}\left( w\right) $ where $w=xt^{\frac{1}{\beta
-1}}$. The reduced system is%
\begin{eqnarray}
\left( x+\left( \beta -1\right) \bar{u}\right) \bar{\rho}_{,x}+2\beta \bar{%
\rho}+\left( \beta -1\right) \bar{\rho}\bar{u}_{,x} &=&0, \\
\left( x+\left( \beta -1\right) \bar{u}\right) p_{,x}+\left( \beta -1\right)
p\bar{u}_{,x} &=&0, \\
x\bar{u}\bar{\rho}_{x}+x\bar{\rho}\bar{u}_{,x}+\beta \bar{u}\bar{\rho}%
+\left( \beta -1\right) \left( \bar{\rho}\bar{u}^{2}+p\right) _{,x} &=&0.
\end{eqnarray}%
For $\beta =1$, the closed-form solution of the later system is $p\left(
w\right) =p_{0},~\bar{\rho}=\rho _{0}w^{-2}$ and $\bar{u}=u_{0}w$.

\subsubsection{Reduction with $X_{1}+\protect\beta X_{6}$}

The similarity transformation which follows from the symmetry vector $%
X_{1}+\beta X_{6}$ is $\rho =e^{\beta t}\bar{\rho}\left( x\right)
,~p=e^{\beta t}\bar{p}\left( x\right) ,~u=\bar{u}\left( x\right) $, where by
replacing in (\ref{tp.32})-(\ref{tp.34}) we find
\begin{equation}
\bar{\rho}=\rho _{0}v_{,x}~,~\bar{p}=p_{0}v_{,x},~\bar{u}=\frac{v_{0}-\beta v%
}{v_{,x}}
\end{equation}%
where $v\left( x\right) $ is a solution of the second-order differential
equation%
\begin{equation}
v_{,xx}\left( p_{0}\left( v_{,x}\right) ^{2}-\rho _{0}\left( \beta
v-v_{0}\right) ^{2}\right) +\beta \rho _{0}v_{,x}^{2}\left( \beta
v-v_{0}\right) =0\text{.}
\end{equation}%
which can be solved by quadratures.

\subsubsection{Reduction with $X_{2}+\protect\beta X_{6}$}

From the symmetry vector $X_{2}+\beta X_{6}$ it follows $\rho =e^{\beta x}%
\bar{\rho}\left( t\right) ,~p=e^{\beta x}\bar{p}\left( t\right) $ and $%
u=u\left( t\right) $, where by replacing in the system (\ref{tp.32})-(\ref%
{tp.34}) it follows%
\begin{eqnarray}
u\left( x\right) &=&u_{1}x+u_{0}, \\
\bar{p}\left( x\right) &=&p_{0}\exp \left( -\beta \left( \frac{1}{2}%
u_{1}x^{2}+u_{0}x\right) \right) , \\
\bar{\rho} &=&-\frac{\beta p_{0}}{u_{1}}\exp \left( -\beta \left( \frac{1}{2}%
u_{1}x^{2}+u_{0}x\right) \right) .
\end{eqnarray}

\subsubsection{Reduction with $X_{3}+\protect\beta X_{6}$}

The Lie symmetry vector $X_{3}+\beta X_{6}$ provides $\rho =t^{\beta }\bar{%
\rho}\left( w\right) ,~p=t^{\beta }\bar{p}\left( w\right) ,~u=u\left(
w\right) $ where $w=\frac{x}{t}$. Hence, from the system (\ref{tp.32})-(\ref%
{tp.34}) we find
\begin{equation}
\bar{\rho}=\rho _{0}\exp \left( -\int \frac{\beta +u_{,w}}{u-w}dw\right) ~,~%
\bar{p}=p_{0}\exp \left( -\int \frac{\beta +u_{,w}}{u-w}dw\right) ,
\end{equation}%
and $u\left( w\right) $ is given by the algebraic equation%
\begin{equation}
\arctan h\left( \frac{e^{\frac{p_{0}}{2}\beta }\left( u\left( w\right)
-w\right) }{\sqrt{\beta +1}}\right) \beta +\sqrt{\left( \beta +1\right) e}%
\left( u_{0}-u\left( w\right) \right) =0.
\end{equation}

In the special case where $\beta =-1\,,$ function $u\left( w\right) $ is
expressed as follows%
\begin{equation}
u_{\pm }\left( w\right) =\frac{1}{2}\left( \left( w-w_{0}\right) \pm \sqrt{%
\left( w-w_{0}\right) ^{2}-4p_{0}}\right) .
\end{equation}

\subsubsection{Reduction with $X_{4}+\protect\beta X_{6}$}

The similarity transformation which correspond to the symmetry vector $%
X_{4}+\beta X_{6}$ is $\rho =e^{\beta \frac{x}{t}}\bar{\rho}\left( t\right)
,~p=e^{\beta \frac{x}{t}}\bar{p}\left( t\right) ,~u=\frac{x}{t}+v\left(
t\right) $. Therefore by replacing in the original system (\ref{tp.32})-(\ref%
{tp.34}) the closed-form solution it follows%
\begin{eqnarray}
v\left( t\right) &=&\frac{u_{1}}{t}+u_{0}~,~ \\
\bar{p}\left( t\right) &=&t^{-1-u_{0}\beta }\exp \left( \frac{u_{1}\beta }{t}%
\right) ~,~ \\
\bar{\rho}\left( t\right) &=&-\frac{\beta p_{0}}{u_{0}}x^{-1-u_{0}\beta
}\exp \left( \frac{u_{1}\beta }{t}\right) .
\end{eqnarray}

\subsubsection{Reduction with $X_{5}+\protect\beta X_{6}$}

From the Lie symmetry $X_{5}+\beta X_{6}$ we calculate $\rho =t^{2-\beta }%
\bar{\rho}\left( x\right) ,~p=t^{-\beta }\bar{p}\left( x\right) $ and $%
u=t^{-1}\bar{u}\left( x\right) $. Thus, from (\ref{tp.32})-(\ref{tp.34}) we
calculate%
\begin{equation}
\bar{p}\left( x\right) =p_{0}\exp \left( -\int \frac{\bar{u}_{,x}-\beta }{%
\bar{u}}dx\right) ~,~\bar{\rho}\left( x\right) =\frac{\bar{u}_{,x}\bar{p}%
-\beta \bar{p}}{\bar{u}_{,x}\bar{u}^{2}-\bar{u}^{2}}.
\end{equation}%
while $\bar{u}\left( x\right) $ is expressed in quadratures, that is
\begin{equation*}
\int_{0}^{\bar{u}\left( x\right) }\left( e^{r\left( s\right) }+\beta \right)
^{-1}ds=x-x_{0},
\end{equation*}%
where $r\left( s\right) $ is a solution of the algebraic equation%
\begin{eqnarray}
0 &=&e^{r\left( s\right) }\beta \left( Z+p_{0}\right) +\left( \beta
-1\right) \left( r\left( s\right) \beta +p_{0}\beta +1\right)  \notag \\
&&-\beta \left( e^{r\left( s\right) }+\beta -1\right) \ln \left( e^{r\left(
s\right) }+\beta -1\right) -2\left( \beta -1\right) \left( e^{r\left(
s\right) }+\beta -1\right) \ln s.
\end{eqnarray}

In the special case where $\beta =1$ the closed-form solution follow%
\begin{equation}
\bar{\rho}=\frac{\rho _{0}}{\left( x-x_{0}\right) ^{2}}~,~\bar{u}=x-x_{0}~,~%
\bar{p}=\rho _{0}.
\end{equation}

\subsubsection{Reduction with $X_{2}+\protect\beta X_{5}+\protect\delta %
X_{6} $}

The similarity transformation which correspond to the vector field $%
X_{2}+\beta X_{5}+\delta X_{6}$ is $\rho =t^{2-\frac{\delta }{\beta }}\bar{%
\rho}\left( w\right) ,~p=t^{-\frac{\delta }{\beta }}\bar{p}\left( w\right)
,~u=t^{-1}\bar{u}\left( w\right) $, where $w=x+\frac{1}{\beta }\ln t$. The
reduced system is
\begin{eqnarray}
\bar{\rho}_{,w}+\left( 2\beta -\delta \right) \bar{\rho}+\beta \left( \bar{%
\rho}\bar{u}\right) _{,w} &=&0, \\
\bar{p}_{,w}-\delta \bar{p}+\beta \left( \bar{p}\bar{u}\right) _{,w} &=&0, \\
\left( \bar{\rho}\bar{u}\right) _{,w}+\left( \beta -\delta \right) \bar{\rho}%
\bar{u}+\beta \left( \bar{\rho}\bar{u}^{2}+\bar{p}\right) _{,w} &=&0,
\end{eqnarray}%
which can be solved by quadratures.

In the limit where $\beta =\frac{\delta }{2}$, the solution is
\begin{equation}
\bar{\rho}=\frac{\rho _{0}}{2+\delta u}~,~\bar{p}\left( x\right) =\rho _{0}%
\frac{\left( \delta \bar{u}+2\right) \bar{u}_{,w}-\delta \bar{u}}{\delta
^{2}\left( \bar{u}-2\right) }
\end{equation}%
where $\bar{u}\left( w\right) $ is given by the first-order differential
equation
\begin{equation}
\frac{df\left( \bar{u}\right) }{d\bar{u}}f-\frac{2\left( f\left( \bar{u}%
\right) -2\right) \left( \left( \bar{u}\delta +2\right) f\left( \bar{u}%
\right) -\left( 2f\left( \bar{u}\right) -1\right) \delta \bar{u}-3f\left(
\bar{u}\right) \right) }{\left( \bar{u}\delta +4\right) \left( \bar{u}\delta
+2\right) },~f\left( \bar{u}\right) =\bar{u}_{,w}.
\end{equation}%
~

\subsubsection{Reduction with $X_{3}+\protect\beta X_{5}+\protect\delta %
X_{6} $}

The symmetry vector $X_{3}+\beta X_{5}+\delta X_{6}$ provides the similarity
transformation $\rho =t^{\frac{2\beta -\delta }{\beta -1}}\bar{\rho}\left(
w\right) ,~p=t^{-\frac{\delta }{\beta -1}}\bar{p}\left( w\right) ,~u=t^{-%
\frac{\beta }{\beta -1}}\bar{u}\left( w\right) $ with $w=xt^{\frac{1}{\beta
-1}}$. Therefore, the reduced system is
\begin{eqnarray}
\bar{\rho}_{,w}+\left( 2\beta -\delta \right) \bar{\rho}+\left( \beta
-1\right) \left( \bar{\rho}\bar{u}\right) _{,w} &=&0, \\
\bar{p}_{,w}-\delta \bar{p}+\left( \beta -1\right) \left( \bar{p}\bar{u}%
\right) _{,w} &=&0, \\
\left( \bar{\rho}\bar{u}\right) _{,w}+\left( \beta -\delta \right) \bar{\rho}%
\bar{u}+\left( \beta -1\right) \left( \bar{\rho}\bar{u}^{2}+\bar{p}\right)
_{,w} &=&0.
\end{eqnarray}%
which can be integrated by quadratures.

When $\beta =1$ and $\delta =-1$ the similarity transformation is $\rho
=x^{-3}\bar{\rho}\left( t\right) ,~p=\bar{p}\left( t\right) $ and $u=x\bar{u}%
\left( t\right) $, where now the exact solution is%
\begin{equation*}
\bar{\rho}\left( x\right) =-\frac{\delta \bar{u}\left( t\right) }{\bar{p}%
^{2}+\bar{p}_{,t}}~,~\bar{u}=u_{0}e^{-\left( \delta +1\right) \int \bar{p}%
dt},
\end{equation*}%
where%
\begin{equation}
\int^{p\left( t\right) }\left( e^{W\left( -\frac{s}{e^{p_{0}}}+p_{0}\right)
}-s^{2}\right) ds=t-t_{0}
\end{equation}%
in which $W\left( t\right) $ is the Lambert function.

\section{Conclusions}

\label{sec4}

In this work we considered a drift-flux two-phase fluids model without
gravitational and wall friction forces where the equation of state parameter
for the fluids is that of a polytropic gas. The system of three hyperbolic
differential equations studied with the use of Lie's theory. In particular
we investigate the algebraic properties of the two-phase fluids model by
calculate the one-parameter point transformations in which the dynamical
system is invariant.

We found that in the general scenario where the polytropic exponents $\gamma
_{1},~\gamma _{2}$ of the two fluids are arbitrary the admitted Lie
symmetries from a algebra of dimension fourth, while when in the special
case where the polytropic exponents are equal, that is, $\gamma _{1}=\gamma
_{2}$ the admitted Lie symmetries form a sixth dimensional Lie algebra. That
result is different from that previously found in the literature where it
was found that a sixth dimensional Lie algebra it is admitted only when $%
\gamma _{1},~\gamma _{2}$ are equal with one, that is, $\gamma _{1}=\gamma
_{2}=1$ \cite{bir1}.

Thus, the missed symmetry vector by the previous study where $\gamma
_{1},~\gamma _{2}$ are equal it is important for the determination of new
similarity transformations. Indeed, because the symmetry vector is defined
only in the two-dimensional space of the dependent variables and it commutes
with the rest symmetries it provides a large number of independent
similarity transformations which lead to similarity solutions which can not
connect through an Adjoint transformation.

Furthermore, the one-dimensional optimal systems were determined for all the
cases which followed by the classification of the Lie symmetries, the
knowledge of the one-dimensional optimal system is essential because all the
unique similarity solutions can be classified. The results of this work
includes and new similarity solutions for the two-phase fluids model of our
study.

{In this work we have not studied the initial value problem for the
two-phase fluids model. Indeed, physical problems are defined with a set of
boundary and initial conditions. Consequently, not all similarity solutions
found in this work will satisfy the initial value problems for all the
physical states. However, we were able to find all the possible similarity
solutions, where these can be constrained according to the initial
conditions. Such analysis extends the scopus of this work and will be
investigated in a future study.}

This study contributes on the subject of the study of the algebraic
properties of hyperbolic equations in fluid dynamics. From the result of
this work it is clear that Lie symmetries play an important role on the
determination of exact solutions in the two-phase fluids models. Although in
this work we studied the simplest drift-flux two-phase fluids model, in the
future we plan to extend our analysis in more general models.

\bigskip

\textbf{Conflict of interest: }This work does not have any conflicts of
interest

\textbf{Funding information:} There are no funders to report for this
submission

\end{document}